\documentstyle[aaspptwo]{article}

\def \etal {{\it et al.}}

\def\gtorder{\mathrel{\raise.3ex\hbox{$>$}\mkern-14mu
             \lower0.6ex\hbox{$\sim$}}}
\def\ltorder{\mathrel{\raise.3ex\hbox{$<$}\mkern-14mu
             \lower0.6ex\hbox{$\sim$}}}
\def \lsim {\ltorder}
\def \gsim {\gtorder}
\def \hide#1{}
\def \half {{1\over 2}}

\def \rms {{\rm rms}}

\def \OmegaGW {\Omega_{\rm GW}}
\def \Hz {{\rm Hz}}

\def \GW {{\rm GW}}

\def \bn {{\bf n}}
\def \bm {{\bf m}}
\def \bb {{\bf b}}
\def \bd {{\bf d}}
\def \bT {{\bf T}}
\def \br {{\bf r}}
\def \bk {{\bf k}}
\def \bH {{\bf H}}

\def \btheta {{\bf \theta}}

\def \cc {{\rm c.c.}}

\def \d	{\partial}
\def \cosphi {{\cos \varphi }}
\def \sinphi {{\sin \varphi }}
\def \cossquphi {{\cos^2 \varphi }}
\def \sinsquphi {{\sin^2 \varphi }}

\begin{document}

\title{Bending of Light by Gravity Waves}
\author{Nick Kaiser and Andrew Jaffe}
\affil{
Canadian Institute for Theoretical Astrophysics \\
60 St George St., Toronto M5S 3H8 \\
e-mail: {\tt kaiser@cita.utoronto.ca} \quad {\tt jaffe@cita.utoronto.ca}
}

\begin{abstract}
We describe the statistical properties of light rays propagating
though a random sea of gravity waves and compare with
the case for scalar metric perturbations from density inhomogeneities. 
For scalar fluctuations the deflection angle
grows as the square-root of the path length $D$ in the manner of
a random walk, and the rms displacement of a ray from the unperturbed
trajectory grows as $D^{3/2}$. For gravity waves the situation
is very different.  The mean square deflection angle remains finite
and is dominated by the effect of the metric fluctuations at the
ends of the ray, and the mean square displacement grows
only as the logarithm of the path length. In terms of power 
spectra, the displacement for scalar perturbations has
$P(k) \propto 1/ k^4$ while for gravity waves the trajectories
of photons have $P(k) \propto 1/k$ which is a scale-invariant
or `flicker-noise' process, and departures from rectilinear
motion are suppressed, relative to the scalar case, by a factor
$\sim (\lambda / D)^{3/2}$ where $\lambda$ is the 
characteristic scale of the metric fluctuations and
$D$ is the path length. 
This result casts doubt on the viability of some recent
proposals for detecting or constraining the gravity wave background
by astronomical measurements.
\end{abstract}

\twocolumn

\section{Introduction}

The deflection of light by scalar metric perturbations generated
by density inhomogeneity is well understood: light propagates
through an inhomogeneous universe much as it would through a block of
glass with inhomogeneous refractive index $n = 1 - 2 \phi = 1 + h$,
and this provides
important observational constraints on the distribution of dark matter 
and associated metric fluctuations $h$ on
a very wide range of scales.  In $\mu$-lensing (angular
separations $\theta \sim 10^{-6}$arcsec (Paczynski, 1986)) 
one is probing metric
fluctuations $h \sim \theta \sim 10^{-11}$, and with sufficiently
compact and distant sources this can be pushed to much weaker
levels: For femto-lensing of cosmologically
distant gamma-ray bursters by comet mass objects (Gould, 1992) the
sensitivity is $h \sim 10^{-20}$ localized in a region
of only $\sim 10^8$cm in size. 
At the other extreme we have weak lensing by large-scale structure
(Blandford \etal, 1991, Miralda-Escude 1991, Kaiser, 1992),
where the shear $\gamma$ and fractional amplification $\delta A$ are on the order of
$\sim D \int dr \nabla^2 \Phi$ where $D$ is the path length, and
give rms shear and amplification
$\gamma, \delta A  \sim h (D / \lambda)^{3/2}$
so the dimensionless shear is larger than the dimensionless
metric fluctuation by a large factor (being the combination of a `lever-arm'
term $D / \lambda$ and a `root-N' factor due to $N \sim D / \lambda$
independent structures along the line of sight adding in quadrature).

It is natural to ask whether one can place similar
constraints on `tensor' metric fluctuations --- 
i.e.~gravitational radiation ---
particularly since the most popular theories for structure formation
such as inflation 
and topological defects  make fairly specific predictions
for the gravity wave background;
(Abbott and Wise, 1984; Battye, \etal, 1996, see Allen, 1996 for
a recent review).   For large-scale structure, the
current amplitude of tensor modes is very small compared to the
scalar fluctuations, but on
small-scales the tensor fluctuations are expected to dominate.
To take an example, popular inflationary models predict
$\OmegaGW \sim h^2 \omega^2 / H_0^2 \sim 10^{-12}$ on small scales;
those which re-entered the horizon in the radiation era, corresponding
to rms metric fluctuations $h \sim 10^{-6} H_0 / \omega$, where
$\omega$ is the frequency of the waves and $H_0$ is the
Hubble parameter. Naively applying
the weak lensing formula above (with $D \sim 1/ H$)
one finds that while the metric fluctuations decrease with
increasing spatial frequency, the amplification and
shear increase as $\sqrt{\omega}$, and reach unity at
$\omega \sim 10^{12} H \sim 10^{-5} \Hz$.  Scalar metric fluctuations of
this amplitude would certainly yield very interesting
effects since $\delta A \sim 1$ signals the onset of
multipath propagation. This would cause scintillation
and observable time variation of distant sources fluxes
(provided the sources have angular size
$\theta \lsim 10^{-12} \sim 10^{-7}$arcsec), and there would also
be potentially interesting interference effects at radio wavelengths.  
However, it is not at all clear that tensor metric perturbations
deflect light in the same way as scalar perturbations. For one thing
the perturbations are moving, and also the waves are transverse
as compared to scalar modes which cause a longitudinal deflection.
As we shall see, both factors have a profound influence
on the character of photon trajectories.

This subject has a long but somewhat confusing history.  
Some analyses have assumed that gravity waves
behave much like scalar perturbations, but with rapid time variation:
Bergman (1971) discussed scintillation from
multi-path propagation. It has been suggested that
gravity waves could cause a time-varying lateral displacement
of the network of $\mu$-lensing induced
caustics in lenses such as 0957+561 (though Linder (1987; cited in
Linder 1988) has suggested that the effect is rather weak,
corresponding to an observer-caustic velocity
of order $\sim h$)  and Fakir, 1994b has discussed the analogous effect
for caustics produced by ISM refractive index inhomogeneity, see below. 
Marleau and Starkman (1996) have discussed
image broadening by gravity wave induced `seeing'.  
Linder (1988) has calculated the apparent clustering of galaxies induced
by amplification by gravity waves.   He finds
an angular correlation function $w(\theta) \sim
D^2 \omega^2 h^2$, which is what one would expect if the rms
amplification were on the order of $\delta A \sim
h D / \lambda$.  This is somewhat different from the
case for scalar metric fluctuations where one finds
$\delta A \sim h (D/\lambda)^{3/2}$, so according to Linder the
rms amplification grows with increasing path length, but more slowly
than would be the case for scalar perturbations.
However, Zipoy and Bertotti (1968) have argued that to first
order in the gravity wave amplitude there is no amplification,
which seems to contradict this.

Braginsky \etal (1990) have calculated
the phase shift of an initially planar EM wave propagating 
through a stochastic GW background.
They find (their equation 4.1) a 
mean square phase shift for receivers separated by baseline $L$
and receiving EM waves of frequency $\omega_\gamma \gg L$ 
\begin{equation}
\langle (\delta \phi)^2 \rangle \sim \omega_\gamma^2 L^2 
(h_{\rm em}^2 + h_{\rm obs}^2)
\end{equation}
This would imply a rms deflection of the light rays
$\delta \theta = \delta \phi / (\omega_\gamma L)$
on the order
of the rms metric fluctuation $h$, but with contributions from the
observer and source adding in quadrature.
A key point here is that the phase shift or ray deflection does not
increase with path length,
in stark contrast with the result for scalar perturbations
where the transverse photon momentum performs a random
walk and $\delta \theta \sim h \sqrt{D/\lambda}$.
The appearance of a term involving $h_{\rm em}$ 
raises the interesting possibility that
one might see relative motions of
sources which appear close together on the sky, and Bar-Kana (1996)
in an independent analysis has obtained a similar result: 
$\dot{\delta \theta}  \sim \omega h$, and
has suggested that with VLBI limits on relative motions
of distant sources this could provide
useful limits on $\OmegaGW$.

The works cited above invoke a statistically homogeneous 
and isotropic stochastic background
of gravity waves.  There have also been analyses of the
deflection of rays passing by a point source of gravity waves such
as a binary star system.  Fakir 1994b has considered the case
where light rays pass close to a binary GW source (impact
parameter $b \sim \lambda_\GW$) and
finds a deflection of order $h(b)$. Durrer (1994) has made a similar
analysis and finds $\delta \theta \sim h(b)$ in general.

In an attempt to clarify the situation we will analyze
the deflection of light rays propagating through
a stochastic background of gravity waves by means of the
geodesic equation.  We recover the result of Braginsky
\etal, but we show that the term involving the metric
fluctuation at the observer does not give rise to observable image
motions.
We find that relative motions of distant sources are
$\dot{\delta \theta} \sim (h/D) \sqrt{\ln(D/\lambda)}$,
much smaller than found by Bar-Kana, and consequently
VLBI observations of relative proper motion cannot
usefully constrain $\OmegaGW$.
We find that the dominant source of proper motions
is the gravity wave at the location of the observer
(which gives rise to a characteristic distortion of the sky as
discussed by Fakir (1992) and Pyne \etal (1996)).  Aside from this we
find that the light rays are remarkably straight --- large-scale
deviations from rectilinear motion are smaller than the case for scalar
perturbations by a factor $\sim (\lambda / D)^{3/2}$
and take the form of a 1-dimensional `flicker noise' process
(see Press, 1978 for a review of flicker noise processes).
For gravity waves, the displacement of a ray grows only
logarithmically with path length rather than as $D^{3/2}$
for scalar perturbations.
We show that stochastic gravity waves do not appreciably displace
microlensing caustics (in general agreement with Linder (1987)), 
neither do they cause galaxy
clustering (counter to Linder (1988)), nor do they induce
observable distortion or rotation of distant objects.

\section{Light Deflection by a Planar Perturbation.}
\label{sec:geodesic}

We now compute the deflection of light by a single
planar (though not necessarily sinusoidal) metric fluctuation.
As we will work in the linear approximation, the deflection
for the general case can be made by superposition of
planar disturbances.

We consider geodesics in a linearly perturbed Minkowski spacetime,
$g_{\mu\nu}=\eta_{\mu\nu}+h_{\mu\nu}$, with the background metric
\begin{equation}
  ds^2 = \eta_{\mu\nu} dr^\mu dr^\nu = -dt^2 + dx^2 + dy^2 + dz^2
\end{equation}
The generalisation to an expanding FRW cosmology is straightforward,
but our main point (the qualitative difference between scalar and
tensor fluctuations) is adequately illustrated by the simpler
flat-space calculation.

Consider a photon with 4-momentum $p^\nu = n^\nu + b^\nu$
where $n^\nu$ is the constant unperturbed vector $n^\nu=(1,\bn)$
with $\bn \cdot \bn = 1$
and $b^\nu$ is the first order perturbation due to $h_{\mu\nu}$.
We can obtain the linearized geodesic equation 
by varying the action obtained from the Lagrangian
\begin{equation}
  L=g_{\mu\nu}{\dot x}^\mu{\dot x}^\nu
  =(\eta_{\mu\nu}+h_{\mu\nu})p^\mu p^\nu.
\end{equation}
(or by variation of the path length $\delta \int dt \sqrt{L} = 0$)
to obtain
\begin{equation}
\label{eq:geodesic}
    {\dot b}^\alpha = - \eta^{\alpha\beta} (n^\mu {\dot h}_{\beta\mu}
- {1\over2}n^\mu n^\nu \partial_\beta h_{\mu\nu})
\end{equation}
We take the time coordinate as the
affine parameter, so the ``dot'' operator is
$(d/dt)=(n^\mu\partial_\mu)$. Here we are interested in the
displacement $\bd$ of a light ray in the plane perpendicular to
the unperturbed ray, and its time derivative
$\dot \bd = \bb - (\bn \cdot \bb) \bn \equiv \bb_\perp$
which is the first order perturbation to direction of the light ray.


For scalar perturbations we can write $h_{\alpha\beta} =
h \delta_{\alpha\beta}$ with $h = -2 \Phi$, where $\Phi$
is the Newtonian gravitational potential, and we find
for the spatial components of the photon momentum
$\dot b_l = -\dot h n_l + \d_l h$.  
For a static planar perturbation with wave direction
$\bm$ we have $h = f(\bm \cdot \br)$ so $\dot h = n_l m_l f'$,
$\d_l h = m_l f'$, and therefore
\begin{equation}
\label{eq:scalargeodesic}
\ddot \bd = {\bm_\perp \over \bn \cdot \bm} \dot h
\end{equation}
where $\bm_\perp \equiv \bm - (\bn \cdot \bm) \bn$.

To obtain the corresponding equation for gravity waves it is
convenient to temporarily adopt a coordinate frame such that
$\bm = (0,0,1)$ and $\bn = (\sqrt{1 - \mu^2}, 0, \mu)$.
The non-zero components
for a transverse traceless gravitational wave
are then
\begin{equation}
\label{eq:transverse}
\begin{matrix}{
h_{xx} = -h_{yy} = h_+(t - z) \cr 
k_{xy} = h_{yx} = k_\times(t - z)
}
\end{matrix}
\end{equation}
and replacing partial derivatives with total time derivatives
much as above (but now with $h = f(t - \bm \cdot \br)$ so
$\d_l h = - m_l f'$, $ \dot h = (1 - \mu) f'$, so 
$\d_l h = - m_l (1 - \mu)^{-1} \dot h$) we find
\begin{equation}
\label{eq:tensorgeodesiccomponents}
\begin{matrix}{
\dot b_x = -  \sqrt{1 - \mu^2} \dot h_+\cr
\dot b_y = -  \sqrt{1 - \mu^2} \dot h_\times\cr
\dot b_z = - \half (1 + \mu) \dot h_+
}\end{matrix}
\end{equation}
The basis vectors in this frame are given by
$\hat {\bf x} = (\bn - (\bm \cdot \bn) \bm) / \sqrt{1 - \mu^2}$, 
$\hat {\bf y} = \bn \times \bm / \sqrt{1 - \mu^2}$, 
$\hat {\bf z} = \bm$, 
so we can write (\ref{eq:tensorgeodesiccomponents})
in vector form as $\dot \bb = \dot b_x \hat {\bf x} + \ldots$, and projecting
$\dot \bb$ onto the plane perpendicular to $\bn$ we have
\begin{equation}
\label{eq:tensorgeodesic}
\ddot \bd = - \half (1 - \bn \cdot \bm) \bm_\perp \dot h_+
- \bn \times \bm \dot h_\times
\end{equation}
This is a vector equation and is therefore valid in an
arbitrary coordinate frame, though it should be kept in mind that
the bases for the decomposition into polarization states 
are still defined by $\bn$, $\bm$.

The `$+$' component drives a deflection in the plane defined
by $\bn$, $\bm$, and $h_\times$  drives a deflection 
perpendicular to both $\bn$, $\bm$.  Note also that the lengths of the
vector coefficients of $h_+$, $h_\times$ are not equal:
$\half \vert (1 - \bn \cdot \bm) \bm_\perp \vert = 
\half (1 - \mu) \sqrt{1 - \mu^2}$
while $\vert \bn \times \bm \vert = \sqrt{1 - \mu^2}$, so for modes
with $\bm$ close to $\bn$ the coupling to the $h_+$ polarization
is relatively suppressed.

A word is in order on the physical meaning of the coordinate system
we have implicitly adopted here.
Since $h_{0\alpha} = 0$, it follows from 
(\ref{eq:geodesic}), (\ref{eq:transverse}) that observers with 
$\dot r^{\alpha} = (1, 0,0,0)$ initially will 
remain at rest in this coordinate system.  Thus equation (\ref{eq:tensorgeodesic})
describes the deflections of rays relative to a coordinate grid that
can be realized physically by a dust of freely falling test particles.
This is convenient since, for the most part,
the sources and observers we will consider 
below are in free fall, and therefore coincide with our coordinate frame
aside from some uniform unaccelerated peculiar motion.

To summarize, for planar perturbations, the geodesic equation
for the transverse deflection of the ray is
\begin{equation}
\label{unifiedgeodesic}
\begin{matrix}{
	\ddot \bd = \bT(\bm, \bn) \dot h & {\rm scalar} \cr
	\ddot \bd = \bT_a(\bm, \bn) \dot h_a & {\rm tensor}
}\end{matrix}
\end{equation}
where, for tensor perturbations, there is implied summation over
the index $a = +,\times$, and where
\begin{equation}
\begin{matrix}{
	\bT = \bm_\perp / (\bn \cdot \bm) \cr
	{\rm and} \cr
	\bT_+ =  - \half (1 - \bn \cdot \bm) \bm_\perp \cr
	\bT_\times = - \bm \times \bn \cr
}\end{matrix}
\end{equation}
Thus, for both scalar and tensor
perturbations, the
transverse acceleration can be expressed as the total time derivative of the
metric times a simple vector valued function of
the photon direction $\bn$ and the wave direction $\bm$.
We can trivially
integrate $\ddot \bd$ to find
e.g.~the change in the ray direction as the difference in the
metric perturbation between the end points of the ray.
To obtain the deflection for
a general perturbation we need to sum over plane wave
components.  This we will do below, but first we note the
qualitative difference between the functions $\bT$, $\bT_a$;
the former has a pole at $\mu = \bn \cdot \bm = 0$ whereas
$\bT_a$ has none.  Thus for scalar perturbations modes
with $\mu \simeq 0$ (which are nearly transverse to the
line of sight) will have a special significance.  This is
easy to understand physically: For these nearly transverse
scalar modes the photon stays in phase with the perturbation
over extended distances and receives a coherent acceleration
over an extended time --- these are the `resonant modes' in the
language of Braginsky \etal --- and the ray acquires
a deflection proportional to $h/\mu$.   For statistically
isotropic random scalar fluctuations, only a small fraction
of the power is in the nearly resonant modes, yet when we
compute the variance in the deflection we find that these
dominate and give a `random walk' for the rms deflection
$\dot d_\rms \sim \sqrt{D / \lambda} h \gg h$.

Tensor perturbations are qualitatively quite different.
Since the wave crests are moving at $c$ the $\mu = 0$
modes are no longer resonant.  The resonant modes are now
$\mu = 1$, where the photon surfs along with the
disturbance, but these produce no deflection due to
the symmetry of a gravity wave and the deflection
for modes close to $\mu = 1$ is also small.
We can immediately see that
for gravity waves $T_a \le 1$
so the net rms deflection can
be no larger than the rms metric fluctuation, and this immediately rules out the 
possibility that the deflection grows as a random walk.

\section{Deflection by Random Perturbations}
\label{sec:stochastic}

We now explore the statistical properties of rays propagating
through a general random background of metric fluctuations.
We first give a rigorous calculation of the 2-point function
and power spectrum for the transverse deflections of a ray.
We then give a heuristic derivation.
To start, without any loss of generality, 
we decompose an arbitrary metric fluctuation into
plane sinusoidal waves:
\begin{equation}
\label{eq:hdecomposition}
\begin{matrix}{
	h(\br) = \half \int {d^3 k \over (2 \pi)^3} h(\bk) 
	e^{i \bk \cdot \br} + \cc & {\rm scalar} \cr
	h_a(\br, t) = \half\int {d^3 k \over (2 \pi)^3} h_a(\bk) 
	e^{i (\omega t - \bk \cdot \br)} + \cc & {\rm tensor} 
}\end{matrix}
\end{equation}
If we now choose spatial coordinates such that the unperturbed
ray lies along the $z$-axis; $\bn = (0,0,1)$ and set
$\bm = \hat \bk = (\sqrt{1 - \mu^2} \cos \varphi, \sqrt{1 - \mu^2} \sin \varphi, \mu)$
then we can write the geodesic equation as
\begin{equation}
\label{eq:unifiedgeodesic}
\ddot \bd(z) = \dot \bH(z)
\end{equation}
where
\begin{equation}
\label{eq:Hdefinition}
\begin{matrix}{
	\bH(z) = \half \int {d^3 k \over (2 \pi)^3} h(\bk) \bT(\bk)
	e^{i \mu k z} + \cc & {\rm scalar} \cr
	\bH(z) = \half \int {d^3 k \over (2 \pi)^3} h_a(\bk) \bT_a(\bk)
	e^{i (1 - \mu) k z} + \cc & {\rm tensor} 
}\end{matrix}
\end{equation}
and where the vectors $\bT$, $\bT_a$ (which lie in the $x,y$ plane)
have components
\begin{equation}
\label{eq:Tcomponents}
\begin{matrix}{
	\bT = \mu^{-1} \sqrt{1 - \mu^2}
	\left[\begin{matrix}{\cos \varphi \cr \sin \varphi}\end{matrix}\right] \cr
	\bT_+ = - \half (1 - \mu)  \sqrt{1 - \mu^2}
	\left[\begin{matrix}{\cos \varphi \cr \sin \varphi}\end{matrix}\right] \cr
	\bT_\times = - \sqrt{1 - \mu^2}
	\left[\begin{matrix}{\sin \varphi \cr -\cos \varphi}\end{matrix}\right]	
}\end{matrix}
\end{equation}

For a static field $h(\br)$ we can simply take $h(k)$ to be the
fourier transform $h(\bk) = \tilde h(\bk) \equiv \int d^3r h(\br)
e^{-i \bk \cdot \br}$, since this implies $h^*(\bk) = h(-\bk)$ and
(\ref{eq:hdecomposition}) is then just the inverse fourier transform.
If the field $h(\br)$ is statistically homogeneous and isotropic,
so the auto-correlation function for the metric fluctuations
$\xi_h = \langle h(\br') h(\br' + \br)\rangle$ is independent
of $\br'$ and depends only on the modulus of $\br$,
 then we find
\begin{equation}
\label{eq:scalarpower}
\langle h(\bk) h^*(\bk') \rangle = (2 \pi)^3 \delta(\bk - \bk') P(k)
\end{equation}
where $P(k) = \int d^3r \xi_h(r) e^{-i \bk\cdot \br}$ is the power spectrum.
At the two-point level then, fourier modes at distinct $\bk$'s are
uncorrelated. However, when computing the mean square values
for observables using (\ref{eq:hdecomposition}) one must allow for
the symmetry between $h(\bk)$ and $h^*(-\bk)$, which imply
$\langle h(\bk) h(\bk') \rangle = \langle h^*(\bk) h^*(\bk') \rangle
= (2 \pi)^3 \delta(\bk + \bk') P(k)$.

For the dynamic radiation field $h_a(\br, t)$ things are slightly different.
Here the amplitudes for the modes in (\ref{eq:hdecomposition})
are related to the fourier transform of the field by
$h(\bk) = \half (\tilde h^*(\bk) + \tilde h'^*(\bk)/ik)$ and
$h^*(\bk) = \half (\tilde h(\bk) - \tilde h'(\bk)/ik)$, 
where we have suppressed the polarization subscript, and
where $\tilde h(\bk)$ is
the fourier transform of $h$ at $t = 0$, and $\tilde {h'}$ is the
fourier transform of $\d h / \d t$, also at $t = 0$, and we now find
\begin{equation}
\label{eq:tensorpower}
\langle h_a(\bk) h_b^*(\bk') \rangle
= (2 \pi)^3 \delta(\bk - \bk') \delta_{ab} P(k)
\end{equation}
but, in contrast to the static case, there is now no correlation
between $h(\bk)$ and $h(-\bk)$.
In both static and dynamic cases the total
metric field variance is 
$\langle h^2 \rangle = \int dk\;k^2 P(k) / (2 \pi^2)$.  
The $\delta_{ab}$ in (\ref{eq:tensorpower})
is a consequence of the assumed statistical isotropy of the
process.

Equation (\ref{eq:tensorpower}) clearly obtains for gravity waves 
from inflation
where the perturbations were initially zero-point
oscillations of the metric, so distinct polarization
states and wave vectors are completely uncorrelated, and
$h(\br)$ is a gaussian random
field, but they are also valid for a background of radiation
from coalescing binaries or from decay of cosmic strings
(Battye \etal, 1996), provided only that the sources are
randomly distributed in space and in orientation.
We should stress that (\ref{eq:tensorpower}) is valid
even if the metric fluctuations are highly non-gaussian,
as, for instance, in the case of a background consisting
of pulses which rarely overlap,
though for these types of backgrounds there may
be non-trivial higher order correlations.  We shall restrict ourselves
to computing variances of observables, for which (\ref{eq:tensorpower})
provides a full description.

To compute e.g.~proper motion of sources etc.~it is necessary
to supply appropriate boundary conditions and then solve
(\ref{eq:unifiedgeodesic}) to obtain the observable of interest
(whose variance can then be computed as an integral
involving $P(k)$).  
This we will do presently for various interesting observable quantities.  
However, the properties
of the trajectories of photons in transit are fully determined by the
statistical properties of the field $\bH(z)$
which is just some statistically homogeneous 1-dimensional
random process.  To explore this we now compute the spatial
2-point function for the $\bH$ field and its analogue in
fourier space, the power spectrum.
From (\ref{eq:Hdefinition}), (\ref{eq:Tcomponents}), (\ref{eq:tensorpower}) we find
$\langle H_m(0) H_n(z) \rangle = \half \delta_{mn} \xi_H(z)$
with
\begin{equation}
\label{eq:tensorxiH}
\begin{matrix}{
\xi_H(z) = \langle \bH(0) \cdot \bH(z) \rangle \cr
= \half \int\limits_0^\infty
 {d^3k \over (2 \pi)^3} P(k) (T_+^2 + T_\times^2)
e^{i (1-\mu) k z} \cr
= {1 \over 32 \pi^2} \int\limits_0^\infty dk k^2  P(k)  \cr
\quad\quad\quad\quad \times
\int\limits_{-1}^{+1} d\mu (5 -2 \mu - 4\mu^2 + 2 \mu^3 - \mu^4) 
e^{i (1-\mu) k z}
}\end{matrix}
\end{equation}
from which we can readily obtain the power-spectrum for the $\bH$-field
\begin{equation}
\label{eq:tensorPH}
\begin{matrix}{
P_H(k) = \int dz \xi_H(z) e^{ikz} \cr
= {1 \over 16 \pi} 
\int\limits_0^\infty dk' k' P(k') \cr
\times \int\limits_{-1}^1
 d\mu (5 -2 \mu - 4 \mu^2 + 2 \mu^3 - \mu^4) 
\delta(\mu - (1 - k/k')) \cr
= {k \over 16 \pi} \int\limits_{k/2}^\infty dk' P(k') 
(8 -  4 k / k' + 2 (k / k')^2 - (k / k')^3)
}\end{matrix}
\end{equation}

For scalar perturbations we find
from (\ref{eq:Hdefinition}), (\ref{eq:Tcomponents}), (\ref{eq:scalarpower})
\begin{equation}
\label{eq:scalarxiH}
\xi_H(z)  = 
{1 \over 4 \pi^2} \int\limits_0^\infty dk k^2 P(k) 
\int\limits_{-1}^1 d\mu (1 - \mu^2)  \mu^{-2} e^{i\mu k z}
\end{equation}
Interestingly, $\xi_H(z)$ is not well defined for scalar perturbations as
the $\mu$-integral diverges at $\mu \rightarrow 0$.  However,
the `structure function' $S_H(z) \equiv
\langle \vert \bH(0) - \bH(z) \vert^2 \rangle = 2(\xi_H(0) - \xi_H(z))$
is well defined, as is the power spectrum
\begin{equation}
\begin{matrix}{
\label{eq:scalarPH}
P_H(k) = {1 \over 2 \pi} \int\limits_0^\infty dk' k' P(k') 
\int\limits_{-1}^1 d\mu (1 - \mu^2) \mu^{-2} \delta(\mu - k/k') \cr
= { 1 \over 2 \pi k^2} \int\limits_k^\infty dk' k'^3 P(k') (1 - k^2/k'^2)
}\end{matrix}
\end{equation}
though this is divergent in the limit $k \rightarrow 0$.

The power spectra (\ref{eq:tensorPH}, \ref{eq:scalarPH}) are quite revealing.
The field $\bH(z)$ is, aside from a constant determined
by boundary conditions, equal to the direction of the ray,
so $P_{\dot d}(k) = P_H(k)$ and the power spectrum for
the displacement is just $P_{d}(k) = P_H(k) / k^2$.
Similarly, the power spectrum for the transverse
acceleration is $P_{\ddot d}(k) = k^2 P_H(k)$.
We are primarily interested in the deflections and displacements
of the beam over long-path lengths $D \gg \lambda_h$,
where $\lambda_h$ is the characteristic wavelength of the
metric fluctuations.
These are determined by the long-wavelength
behavior of $P_{\dot d}(k)$ at $k \ll k_h$, in which case
we can drop the terms involving $k$ within the integrals
and also relace the lower limit on the $k_h$-integration
by zero and find the leading order term for the displacement
power spectrum
\begin{equation}
\label{eq:asymptoticP}
P_d(k) = \left\{
\begin{matrix}{
k^{-4} \int {dk'\over 2 \pi} k'^3 P(k') \sim
\langle h^2 \rangle k_h / k^4 
& {\rm scalar} \cr
k^{-1} \int {dk'\over 2 \pi} P(k') \sim
\langle h^2 \rangle / (k_h^2 k)
& {\rm tensor}
}\end{matrix} \right.
\end{equation}
Thus the effect of both types of metric fluctuations is to produce
a universal scaling law for the power-spectrum of displacements
at $k \ll k_h$, but the form of the scaling is quite different:
Scalar modes give a displacement
which is the double integral of a `white-noise' process: 
$P_d(k) \propto k^{-4}$, and
the mean-square deviation, averaged over a scale $D$ is
$\langle d^2 \rangle_D \sim (k P_{d})_{k = 1/D} \propto D^3$.
Tensor modes, on the other hand, generate a `$1/f$' or
`flicker noise' spectrum $P_d(k) \propto k^{-1}$ with scale-invariant deviation
$\langle d^2 \rangle \sim (k P_{d})_{k = 1/D} 
\sim \lambda_h^2\langle h^2 \rangle \propto D^{0}$.
For a given
scale and amplitude of metric fluctuations the large-scale deviations
from rectilinear motion are suppressed by a factor $\sim (\lambda / D)^{3/2}$
for tensor modes as compared to scalar modes.

Another interesting feature emerges if we consider the
power spectrum for the acceleration at zero frequency.
In the scalar case $P_{\ddot d}(k)$ tends to a constant value
at $k \rightarrow 0$ whereas for tensor fluctuations
$P_{\ddot d}(k)$ being proportional to $k^3$ vanishes
in this limit.  However, the power at zero frequency is the integral of
the corresponding 2-point function:
$P_{\ddot d}(0) = \int dz \xi_{\ddot d}(z)$.
What this is telling us is that for scalar perturbations the
sequence of accelerations suffered by a photon can be 
legitimately modeled as a series of uncorrelated kicks
(as this gives the correct random walk for the
photon direction) whereas 
for tensor fluctuations the accelerations are strongly
anticorrelated, and this again helps us understand why the net deflections
are so much smaller for gravity waves.

Equation (\ref{eq:asymptoticP}) give the universal asymptotic
scaling laws for the low-frequency power spectrum of photon trajectories
propagating through metric fluctuations of relatively higher spatial frequency.
They are the main result of this paper and show immediately that
tensor metric fluctuations are extremely inefficient at deflecting
light.  

We can give a somewhat simpler derivation of these scaling laws.
The mean-square systematic deflection of a ray over path length $D$, is
proportional to the power in spatial modes which project to
low spatial frequency $\sim 1/D$ along the line of sight, times
the square of the amplitude factor $T$.    
For scalar perturbations, if we compute $\langle \dot \bd^2 \rangle_D
\sim (k P_{\dot d}(k))_{k\sim 1/D}$ we are only sensitive to modes
which have $\mu \sim 1/(k_h D)$, where $k_h$ is the 
spatial frequency of the metric fluctuation.
These are a fraction $\sim \mu$ of all modes, and
have an amplitude factor $T \sim 1/\mu$ so
$\langle \dot \bd^2 \rangle_D \sim \langle h^2 \rangle (k_h D)^{-1} T^2
\sim k_h D \langle h^2 \rangle$ and hence $\langle \bd^2 \rangle_D 
\sim D^2 \langle \dot \bd^2 \rangle_D \sim k_h D^3 \langle h^2 \rangle$.
For tensor fluctuations, the modes which project to line-of-sight
wave-vector $k \sim 1/D$ are those with $(1 - \mu) \sim 1 / k_\GW D$.
These again are a fraction $\sim 1 / k_\GW D$ of all modes, but they have
$T_+ \sim (k_\GW D)^{-3/2}$, $T_\times \sim (k_\GW D)^{-1/2}$
and consequently, for $k_\GW D \gg 1 $,
the $\times$-polarization components dominate,
$\langle \dot \bd^2 \rangle_D \sim \langle h^2 \rangle (k_\GW D)^{-1} T_\times^2
\sim \langle h^2 \rangle / (k_\GW D)^2$ and therefore
$\langle \bd^2 \rangle_D 
\sim D^2 \langle \dot \bd^2 \rangle_D \sim k_\GW^{-2} \langle h^2 \rangle$
which is independent of scale $D$.

We will now work
through some calculations of what are, in principle, observable
quantities. From the order-of-magnitude estimates above
this is largely an academic exercise, but will reveal
some subtleties in handling boundary conditions.

\section{Applications}

We will now apply the results obtained above to compute 
the deflection of a collimated beam (\S\ref{sec:collimatedbeam});
proper motions of sources (\S\ref{sec:propermotions});
deflection of caustics (\S\ref{sec:wigglingcaustics});
and amplification, distortion and rotation of images of distant galaxies
(\S\ref{sec:amplificationandshear})

\subsection{Deflection of a Collimated Beam}
\label{sec:collimatedbeam}

Let us calculate the change in the direction $\dot \bd$
for a ray which propagates from $z_0$ to $z_1 = z_0$.  Clearly, from
(\ref{eq:unifiedgeodesic}) we have
\begin{equation}
\label{eq:tensordeltaddot}
\delta \dot \bd = \dot \bd(z_1) - \dot \bd(z_0) = \bH(z_1) - \bH(z_0) 
\end{equation}
so
\begin{equation}
\begin{matrix}{
\langle |\delta \dot \bd|^2 \rangle = 2 (\xi_H(0) - \xi_H(D)) \cr
{1 \over 16 \pi^2} \int\limits_0^\infty dk k^2 P(k) \cr
\times
\int\limits_{-1}^{+1} d\mu (5 -2 \mu - 4\mu^2 + 2 \mu^3 - \mu^4) 
(1 - e^{i (1-\mu) k D})
}\end{matrix}
\end{equation}
where $D \equiv z_1 - z_0$ is the path length.
For $k D \gg 1$ we can neglect the oscillatory term in the
$\mu$-integral --- i.e.~$\xi_H(D) \ll \xi_H(0)$ --- and
so the integral is elementary and we have
\begin{equation}
\label{eq:tensordeflection}
\langle |\delta \dot \bd|^2 \rangle = {13 \over 15} \langle h^2 \rangle
\end{equation}
This is in accord with the order of magnitude
estimate of Braginsky \etal.
Clearly, the directional change does not
grow with increasing path length. Compare this with
the analogous situation for scalar fluctuations where we have
\begin{equation}
\langle |\delta \dot \bd|^2 \rangle  = 
{1 \over 4 \pi^2} \int\limits_0^\infty dk k^2 P(k) 
\int\limits_{-1}^1 d\mu {(1 - \mu^2) \over \mu^2}
(1 - e^{i\mu k D})
\end{equation}
for fixed $k$ and for $D \gg 1/k$ the $\mu$ integration
is now dominated by $\mu \sim 1 / (kD)$ (these are the
`resonant modes' where the photon sees a phase change over the
path length on the order of unity) and we have
\begin{equation}
\label{eq:scalardeflection}
\begin{matrix}{
\langle |\delta \dot \bd|^2 \rangle  \simeq
{D \over 4 \pi^2} \int\limits_0^\infty dk k^3 P(k) 
\int\limits_{-\infty}^\infty dy y^{-2} (1 - e^{iy}) \cr
\sim D k_* \langle h^2 \rangle
}\end{matrix}
\end{equation}
where we have defined the characteristic wave-number
\begin{equation}
k_* \equiv {\int dk k^3 P(k) \over \int dk k^2 P(k)}
\end{equation}
This is very different from (\ref{eq:tensordeflection}) 
and the variance in the deflection
angle now grows linearly with path length as expected for
a random walk.

Note that the ratio of (\ref{eq:scalardeflection}) to (\ref{eq:tensordeflection})
is not $\sim D^3 / \lambda_h^3$, as one might have anticipated from the
discussion of the previous section.   This is because  (\ref{eq:tensordeflection})
is dominated by `surface terms', rather than low-frequency power with $k_h \sim 1/ D$.

\subsection{Proper Motions}
\label{sec:propermotions}

In the preceding section we
found that the mean square change in direction of a ray
propagating through random gravity waves is essentially the sum of the
mean square metric fluctuations at the end points.  Does this mean
that two sources close together on the sky but at
different distances would show a relative proper motion on the sky
with amplitude $\delta \dot \theta \sim h \omega$?  If so, 
then with VLBI observations
one could constrain $\OmegaGW$ as described by Bar-Kana, 1996.
Unfortunately this is not the case.  The foregoing correctly
gives the final direction of a collimated beam which initially
has $\dot \bd = 0$.
Physically, one could realize this with a pair of neighboring
freely falling observers, one of whom shines a flashlight 
at the other who is holding a plate with a pin-hole
in the centre which selects a particular beam.  This is rather
like a lighthouse, where a beam suffers a deflection at the
source, but where distant observers do not see a moving source,
rather they see a flashing source (and whenever they actually see
the source, they see it at the same location).

Clearly, to compute how real astrophysical sources
--- which are generally at best poorly collimated ---  would move
we need to `solve the lens equation' and find the final
direction for a ray which leaves the source at $z_0$ and
actually arrives at the location of the observer $\bd(z_1) = 0$.
This is readily done. Integrating (\ref{eq:unifiedgeodesic}) once more we find
\begin{equation}
\bd(z) = (z - z_0) \left[ \dot \bd(z_0) - \bH(z_0) \right] +
\int_{z_0}^z dz \bH(z)
\end{equation}
so the beam which reaches a detector at $\bd(z_1) = 0$ must set out with
\begin{equation}
\dot \bd(z_0) = \bH(z_0) - {1\over z_1 - z_0} \int_{z_0}^{z_1} dz \bH(z)
\end{equation}
and consequently will have a direction at the observer
\begin{equation}
\label{eq:sourcedeflection}
\dot \bd(z_1) = \bH(z_1) - {1\over z_1 - z_0} \int_{z_0}^{z_1} dz \bH(z)
\end{equation}
from which we see that the effect of the metric fluctuation
at the source enter only as a minor contribution to the
integral of the metric fluctuations along the line of sight.

Clearly, for a random fluctuating $\bH(z)$ and distant
sources the second term will
be sub-dominant.  However, the first term depends only on the
metric fluctuations at the point of observation, and is a slowly varying
function of the angle between the wave-direction and the
line-of sight.
This term causes a characteristic distortion of the sky which
is the same for all distant ($D \gg \lambda$) sources.
This is the effect discussed by Fakir (1992), Pyne \etal (1996).  
One has to be careful
in how to interpret this term as we have derived it.  The direction
here is defined as relative to our coordinate system which is tied
to freely-falling test particles.  Thus $\dot \bd$ here would correctly
describe the apparent motion of a distant source relative
to a nearby ($D \ll \lambda$) freely-falling foreground reference source,
but would not directly describe the change in relative angles between
distant sources as measured with a sextant.  To compute
the latter we need to calculate how our nearby reference source
would appear to move in the physical reference system.
This is straightforward, and
gives a somewhat different pattern of displacements
(and we recover their equation 51), but does
not change the general result that the angular deflection is on
the order of $h$ and is slowly varying across the sky (and so would
not be detectable by measuring relative proper motions of
neighboring sources with VLBI).

Let us now estimate the size of the second term in (\ref{eq:sourcedeflection}), 
which depends on the metric fluctuations along the line of sight to
the source, and which would therefore be expected to give rise to
relative motions of sources which appear close together on the
sky but lie at different distances.
The variance of the integral $\int dz \bH(z)$ is
\begin{equation}
\label{eq:Hintegralvariance}
\left\langle \left(\int_{z_0}^{z_1} dz H(z)\right)^2 \right\rangle
= \int {dk \over 2 \pi} P_H(k) {\sin^2 k D / 2 \over k^2}
\end{equation}
Now from (\ref{eq:tensorPH}) we see that $P_H(k)$ consists of four
terms of the form $k^n \int_{k/2} dk' k'^{1-n} P(k')$ for
$n = 1 \ldots 4$, each of which rises as a power law for
$k \ll k_\GW$ but then falls to zero for $k \gg k_\GW$ as the
lower limit on the integral takes effect.  It is not difficult to
see that for $k_\GW D \gg 1$ the leading contribution to
the variance above comes from the first term $P_H \sim k \int_{k/2} dk' P(k')$
and the integrand in (\ref{eq:Hintegralvariance}) $\sim \int dk' P(k')/kD^2$
and so there is equal contribution to the variance above for
each logarithmic interval in the range $D^{-1} \ll k \ll k_\GW$ (with the
$\sin^2 k D$ factor limiting the divergence at low $k$ and
the lower limit on the $\int dk' P(k')$ integral limiting the divergence
for $k \gsim k_\GW$) and
\begin{equation}
\label{eq:Hintegralvariance2}
\left\langle \left( \int_{z_0}^{z_1} dz H(z)\right)^2 \right\rangle
\simeq \lambda_\GW^2 \ln(D / \lambda_\GW) \langle h^2 \rangle
\end{equation}
which is just what one would anticipate from the general idea that the
photon trajectories are a $P_d \propto 1/k$ or flicker noise
process. To obtain the mean square angular deflection we simply divide
(\ref{eq:Hintegralvariance2}) by the source distance $D^2$.

Thus the net effect of the the perturbations along the line of sight
is to displace the image by an angle $\theta$ on the order of 
$(\lambda / D)  \sqrt{\ln(D/\lambda)} h $ 
which is very small indeed.  Of course, as we do not
know the unperturbed image location this is not directly observable, but its
time derivative $\dot \theta \equiv
\d \theta / \d t$ would be observable as a proper motion
(or as a relative proper motion for two sources which are close together
on the sky but lie at different distances).  It is straightforward to
compute $\langle \dot \theta^2 \rangle$, since the time variation simply
introduces an extra factor $k_h^2$ in e.g.~(\ref{eq:tensorPH}) so the power
spectrum for $\d H / \d t$ is $\sim k \int dk_h k_h^2 P(k_h)$, and we find
\begin{equation}
\langle \dot \theta^2 \rangle^{1/2}
\simeq (h / D) \sqrt{\ln(D / \lambda_\GW)} \sim \OmegaGW^{1/2} H_0 \lambda_\GW / D
\end{equation}
For gravity waves with periods $1/\omega \sim 10$ yr considered by
Bar-Kana (1996), and for cosmologically distant sources
the rms proper motion is $\delta \dot \theta 
\sim 10^{-14} \sqrt{\OmegaGW}$ arcsec/yr which is $\sim 9$ orders
of magnitude smaller than obtained
by Bar-Kana, and would be rather hard to measure.
Similarly, one can estimate the broadening of images due to
gravity waves (which must have periods less than the integration
time $t_{\rm int}$) to be 
$\theta \lsim H_0^2 t_{\rm int}^2 \sqrt{\OmegaGW}$ 
which would be
on the order of $10^{-21} \sqrt{\OmegaGW}$ arcsec for integration times
of a few hours.

We should emphasize that this is only the relative motion of sources.
The first term in (\ref{eq:sourcedeflection}) gives a much larger
motion $\dot \theta \sim \omega h \sim \OmegaGW^{1/2} H_0$, 
but as discussed above this is a 
purely local distortion effect and would require
precise relative astrometry over large angular scales, and would not be
easily accessible with current technology.

For scalar fluctuations, the variances of $\bH$ and $\int dz \bH(z)$ are
both ill defined due to divergent low-$k$ behavior noted above,
but the variance of the combination
in (\ref{eq:sourcedeflection}) is however well defined, and
$\langle \delta \theta^2 \rangle \sim (D / \lambda) 
\langle h^2 \rangle$.  Note the qualitative difference in the
dependence of the angular displacement on source distance; 
for scalar perturbations
the deflection grows without limit, while for tensor fluctuations,
and aside from the local sky distortion offect,
the deflection tends to zero for very distant sources.  

\subsection{Wiggling Caustics}
\label{sec:wigglingcaustics}

We found above (\ref{sec:collimatedbeam})
that a collimated beam can suffer a deflection due to the metric fluctuation
at the
source which leads to a relatively large `lighthouse' effect, but noted that
highly collimated sources were rare in Nature.  However, a notable
exception is the caustic network produced by $\mu$-lensing whose sharp
cusp-like features cause observable time variations of source intensities
as they sweep past an
observer. The same is true of caustics arising from refractive index inhomogeneity
from the ISM.  This raises the interesting possibility that gravity waves,
if they can laterally displace the caustics, might give rise to anomalous time
variations of sources which might be recognizable.
Fakir (1994b)  has studied the deflection of caustics which
pass close ($b \sim \lambda_\GW$) to a binary GW source.  Here we
will compute the effect of a stochastic background.

We can readily compute the relative motion between a caustic network and a
freely falling observer.  Consider a source at $z_0$ and a lens at $z_1$,
and let both of these have $d = 0$.  Now consider a ray which propagates
from the source to the lens. As before, imposing the boundary conditions
$\bd(z_0) = \bd(z_1) = 0$ 
gives the direction
on arrival at the lens of
\begin{equation}
\dot \bd_1 = H_1 - {1 \over D_{01}} \int\limits_{z_0}^{z_1}
dz \bH
\end{equation}
so, as the photon propagates to $z > z_1$ it will have
\begin{equation}
\dot \bd = \dot \bd_1 + \bH - \bH_1 = \bH - {1 \over D_{01}}
\int\limits_{z_1}^{z} dz \bH
\end{equation}
and hence will arrive at $z_2$ with
\begin{equation}
\bd_2 = \int\limits_{z_1}^{z_2} dz \dot \bd
= \int\limits_{z_1}^{z_2} dz \bH
- {D_{12} \over D_{01}} \int\limits_{z_0}^{z_1} dz \bH
\end{equation}
If both $D_{01}$, $D_{12}$ are $\gg \lambda_\GW$ and are of
similar magnitude then $d_2$ is composed of integrals of the same
form as in the previous section and consequently
$\langle \bd_2^2 \rangle^{1/2}  \sim \lambda_\GW h 
\sqrt{\ln(D / \lambda_\GW)}$, which is very small, and
the velocity with which the caustic moves across the
observer plane is $v \sim h \sqrt{\ln(D / \lambda_\GW)}$
which, aside from
the logarithmic factor, agrees with the rough
estimate of Linder (1987).
The exception is if the lens is very close to the source,
where we can get a somewhat larger effect.  For $D_{01} \ll \lambda_\GW$,
$\int\limits_{z_0}^{z_1} dz \bH \simeq D_{01} \bH_0$ and
we get a stronger `lighthouse' effect with $\bd_2 = D_{12} \bH_0$ and
caustic-observer velocity ${\bf v} = \d \bd_2 / \d t \simeq  D_{12} \bH_0 /
\lambda_\GW \sim D_{12} h / \lambda_\GW$. 
Unfortunately, however, the cross-section for $\mu$-lensing (of for formation
of caustics by inhomogeneous ISM electron density) vanishes for
$D_{01} \rightarrow 0$, and the typical caustics have
$D_{01} \simeq D_{12}$ and the effect is very small, and is certainly
much smaller than typical peculiar motions of observers and lenses for
any plausible value of $\OmegaGW$.

\subsection{Source Amplification, Distortion and Rotation}
\label{sec:amplificationandshear}

As a final application, we consider the possibility of amplification,
distortion and rotation of distant sources.  As discussed in the
Introduction, Linder (1988) has suggested that deflection of light
by gravity waves will change the number density of sources on
the sky in an inhomogeneous way and give rise to
apparent galaxy clustering.  For static lenses, 
the change in number density is
directly related to the amplification of the flux density, since
if the lens mapping increases the area (and hence flux, 
as surface brightness is conserved for static lenses) of a source
by some factor $A$, then it will also decrease the 
expected number density of sources
by the same factor.  This is for `standard-candles'.  For real sources we
have to worry about the change in detectability with flux, but in general
we expect the change in surface density of sources to be proportional
to the amplification (with a constant of proportionality known as the
`amplification bias factor' which can be positive or negative depending on the
slope of the number-flux relation).  However, as we also mentioned,
Zipoy and Bertotti (1968) have argued that there is no amplification
due to gravity waves, which seems to contradict this.  To resolve this
we will estimate the expected effecting using our formalism.  We will
also discuss the distortion and rotation of distant sources, which are
not excluded by Zipoy and Bertotti's argument.

This calculation differs from the previous applications in several respects.
First, to calculate the amplification, distortion and rotation of sources in
a single plane
we need to solve the geodesic deviation equation for a pair of
neighboring rays to obtain the mapping between angles at the
observer and distances on the source plane (which we can express
in terms of an amplification matrix $\delta_{lm} + \psi_{lm}$
where the `distortion tensor' $\psi_{lm}$ is linear in the
perturbation amplitude). Second, we then need to average this over
source planes distributed according to the selection function
for the galaxies. Also, we need to include the possibility
of modulation of the surface brightness due to the non-stationarity
of gravity waves.

For scalar perturbations, it is well known that the mean square
amplification is on the order of $(D/\lambda)^3 \langle h^2 \rangle$
(and hence the induced angular correlation function
$w(\theta)$) grow with depth of survey as $D^3$.
Linder (1988) found
$w(\theta) \sim (D/\lambda)^2 \langle h^2 \rangle\propto D^2$.  However,
this would be hard to reconcile with the result of \S\ref{sec:propermotions}
where we  found that the angular displacement of a distant source was
composed of two terms: a large angular scale distortion of the 
sky due to the gravity wave amplitude here and now, which is on the
order of $h$, and is the same for all distant ($D \gg \lambda$) sources,
and a small extra deflection $\delta \theta \sim (\lambda / D)
\sqrt{\ln(D/\lambda)} h$ which depends on the source location,
and tends to zero for very distant sources.  
Now the components of the distortion
tensor $\psi_{lm}$ are the angular derivatives of the deflection
angle.  Thus, the local distortion effect would be expected to produce
$\psi \sim h$, since the coherence scale for this term is on the
order of one radian.  The non-local term due to metric fluctuations
along the line of sight will cause a fluctuating distortion
which may be more accessible to e.g.~galaxy clustering studies.
If one imagined that the coherence scale were $\theta_c \sim
\lambda / D$ i.e.~the angle subtended by one wavelength at the
distance of the source, then one would predict distortion $\psi
\sim \sqrt{\ln(D/\lambda)} h$ which is somewhat larger than the
local sky distortion effect (though still very small compared to
Linder's calculation). In fact, as we shall see, even this overestimates
the true amplitude of the effect.

Consider two neighboring rays which are nearly parallel
to the $z$-axis. Let the separation between the rays be
\begin{equation}
\Delta \br = \Delta \br_0 + \Delta \bd
\end{equation}
where the zeroth order separation obeys 
$\ddot{\Delta \br_0} = 0$, and where
\begin{equation}
\label{eq:geoddev}
\Delta \ddot d_l = \Delta r_{0p} \d_p \ddot d_l =
\Delta r_{0p} \phi_{pl}
\end{equation}
where the latter
equality defines the two-dimensional transverse tidal field tensor
$\phi_{pl}$.
For scalar perturbations, $\phi_{pl} = \d_p \d_l h$, which is a statistically
homogeneous random field with fourier decomposition
\begin{equation}
\label{eq:scalartidedef}
\phi_{pl} = - \half \int {d^3 k \over (2 \pi)^3} k^2 R_{pl}(\bk) 
h(\bk) e^{i\mu k z} + \cc
\end{equation}
where
\begin{equation}
\label{eq:Rlmdefinition}
R_{lm}(\bk) \equiv 
(1 - \mu^2) \left[\begin{matrix}{
\cossquphi & \cosphi \; \sinphi \cr \cosphi \; \sinphi & \sinsquphi
}\end{matrix}\right]
\end{equation}
and where we have used $\d_p \rightarrow ik_p = \sqrt{1 - \mu^2}
\{\cosphi, \sinphi\}$.
For tensor perturbations, we had $\ddot \bd = \bT_a\dot h_a$, or
from (\ref{eq:unifiedgeodesic}), (\ref{eq:Hdefinition})
\begin{equation}
\ddot d_l = \half {d \over dt} \int {d^3 k \over (2 \pi)^3}
T_l(\bk) h(\bk) e^{i(\omega t - \bk \cdot \br)} + \cc
\end{equation}
so the tidal field tensor is
\begin{equation}
\label{eq:tensortidedef}
\begin{matrix}{
\phi_{pl} = \d_p \ddot d_l = -\half \int {d^3 k \over (2 \pi)^3}
T_l(\bk) h(\bk) (1 - \mu) k k_p e^{i(1-\mu)k z} + \cc \cr
= \half \int {d^3 k \over (2 \pi)^3} k^2
R_{alm} h_a(\bk)  e^{i(1 - \mu) k z} + \cc
}\end{matrix}
\end{equation}
where we have defined
\begin{equation}
\begin{matrix}{
R_{+lm} = \half (1 - \mu^2) (1 - \mu)^2 
\left[\begin{matrix}{
\cossquphi & \cosphi \; \sinphi \cr \cosphi \; \sinphi & \sinsquphi
}\end{matrix}\right] \cr
R_{\times lm} = (1 - \mu^2) (1 - \mu) 
\left[\begin{matrix}{
\cosphi\;\sinphi & -\cossquphi \cr \sinsquphi & -\cosphi\;\sinphi
}\end{matrix}\right]
}\end{matrix}
\end{equation}

To obtain the shear tensor for a source plane at some distance
$z_s$ we need to integrate (\ref{eq:geoddev}) back along the
null rays to obtain the
source plane separation $\Delta \br$ for a pair of rays which
arrive at the observer with angular separation $\btheta$.
These rays have zeroth order separation $\Delta \br_0 = 
\btheta z$, so, with boundary conditions $\Delta \bd = \Delta \dot \bd = 0$
at the observer the solution of (\ref{eq:geoddev})
is just the particular integral:
\begin{equation}
\Delta d_l(z_s) = \theta_p \int\limits_0^{z_s} dz
\int\limits_0^{z} dz' z' \phi_{lp}(z')
= \theta_p \int\limits_0^{z_s} dz z (z_s - z) \phi_{lp}(z)
\end{equation}
and hence obtain the mapping between source plane separation
$\Delta \br = \Delta \br_0 + \Delta \bd$ and angle:
\begin{equation}
\Delta r_l(z_s) = \theta_m z_s \left[
\delta_{lm} + \psi_{lm}\right]
\end{equation}
where the distortion tensor is a certain integral along the line of sight
of the tide:
\begin{equation}
\psi_{lm}(z_s) = \int\limits_0^{z_s} dz z (1 - z / z_s) \phi_{lm}(z)
\end{equation}
We are interested in the expectation value for the distortion
taken over the distribution of distances $n(z)$ for the galaxies
whose clustering, shapes etc.~we are measuring, which we can express
as
\begin{equation}
\label{eq:phibardef}
\overline \psi_{lm} = \int\limits_0^\infty d z_s n(z_s) \psi_{lm}(z_s)
= \int\limits_0^\infty dz \phi_{lm}(z) g(z)
\end{equation}
with
\begin{equation}
\label{eq:gdefinition}
g(z) \equiv \int\limits_z^\infty dz' n(z') (z - z^2 / z')
\end{equation}
and where we have taken $n(z)$ to be normalized to $\int dz n(z) = 1$
so, on substituting (\ref{eq:scalartidedef}),
and (\ref{eq:tensortidedef}) in (\ref{eq:phibardef}) we find
\begin{equation}
\overline \psi_{lm} = \left\{
\begin{matrix}{
\half \int {d^3 k \over (2 \pi)^3}
k^2 R_{lm}(\bk) h(\bk) \tilde g(\mu k) + \cc \cr
\half \int {d^3 k \over (2 \pi)^3}
k^2 R_{alm}(\bk) h_a(\bk) \tilde g((1 - \mu) k) + \cc
}\end{matrix}\right.
\end{equation}
for scalar and tensor perturbations respectively,
and where
\begin{equation}
\label{eq:gtildedefinition}
\tilde g(k) \equiv \int dz g(z) e^{ikz}
\end{equation}
We can now compute the variance of the distortion matrix elements.
First though, it is convenient to decompose the 
$2 \times 2$ tensor $\overline \psi_{lm}$ into
a four component entity
\begin{equation}
\label{eq:gammadecomposition}
\Gamma_\alpha = M_{\alpha l m} \overline\psi_{lm}
\end{equation}
with 
\begin{equation}
\begin{matrix}{
M_0 = \left[\begin{matrix}{1 & 0 \cr 0 & 1}\end{matrix}\right] &
M_1 = \left[\begin{matrix}{1 & 0 \cr 0 & -1}\end{matrix}\right] \cr
M_2 = \left[\begin{matrix}{0 & 1 \cr 1 & 0}\end{matrix}\right] &
M_3 = \left[\begin{matrix}{0 & 1 \cr -1 & 0}\end{matrix}\right]
}\end{matrix}
\end{equation}
so $\Gamma_0$ is the convergence, $\Gamma_1$, $\Gamma_2$ are the shear
and $\Gamma_3$ describes rotation of the image. 
Summing over fourier components, we find
\begin{equation}
\label{eq:gammacovariance}
\langle \Gamma_\alpha \Gamma_\beta \rangle = \left\{
\begin{matrix}{
\int {d^3 k \over (2 \pi)^3}
k^4 P(k) R_\alpha(\bk) R_\beta(\bk) | \tilde g(\mu k) |^2 \cr
\half \int {d^3 k \over (2 \pi)^3}
k^4 P(k) R_{a\alpha}(\bk) R_{a\beta} (\bk) | \tilde g((1 - \mu) k) |^2 
}\end{matrix}\right.
\end{equation}
where
\begin{equation}
\label{eq:Ralphadef}
\begin{matrix}{
R_\alpha = (1 - \mu^2) [1, \cos 2 \varphi, \sin 2 \varphi, 0] \cr
R_{+\alpha} = \half (1 - \mu^2) (1 - \mu)^2 [1, \cos 2 \varphi, \sin 2 \varphi, 0] \cr
R_{\times\alpha} = (1 - \mu^2) (1 - \mu)[0, \sin 2 \varphi, \cos 2 \varphi, 1]
}\end{matrix}
\end{equation}
Performing the $\varphi$-integration in (\ref{eq:gammacovariance})
we find that the decomposition
(\ref{eq:gammadecomposition}) results in a covariance matrix
$\langle \Gamma_\alpha \Gamma_\beta \rangle$ which is diagonal.
For scalar perturbations $\Gamma_3$ vanishes --- no image rotation ---
and $ \langle \Gamma_1^2 \rangle =  \langle \Gamma_2^2 \rangle =
\half \langle \Gamma_0^2 \rangle$ with mean square convergence
\begin{equation}
\label{eq:scalarGammaVariance}
\langle \Gamma_0^2 \rangle =  {1 \over 4 \pi^2}
\int\limits_0^\infty dk k^6 P(k) 
\int\limits_{-1}^{+1} d\mu (1 - \mu^2)^2
| \tilde g(\mu k) |^2
\end{equation}
To estimate the value of the $\mu$-integral here we need to know the
asymptotic form of $\tilde g(k)$ for large and small $k$.
One can see on dimensional grounds from (\ref{eq:gdefinition}) 
(\ref{eq:gtildedefinition}) that
$| \tilde g(y) |^2 \sim D^4$ for $y \lsim 1/D$.  To obtain the
high frequency asymptote we use
\begin{equation}
\label{eq:tildeg}
\begin{matrix}{
\tilde g(k) = \int\limits_0^\infty dz e^{i k z}
\int\limits_z^\infty dz' n(z') (z - z^2 / z') \cr
= \int\limits_0^\infty dz' n(z') \int\limits_0^{z'} dz (z - z^2 / z')
e^{i k z} \cr
= - {1\over k^2} \int\limits_0^\infty dz n(z) (1 + e^{ikz})
+ {2i\over k^3} \int\limits_0^\infty dz {n(z) \over z} (1 - e^{ikz})
}\end{matrix}
\end{equation}
In general one can consider $n(z)$ to be  composed of the smoothly
varying expectation value $\overline n(z)$ plus a fluctuating
component due to galaxy clustering.  We will first consider
the smooth term $n = \overline n$.
To obtain the leading order behavior of the two fourier type integrals
at high $k$ we
use the series expansion, obtained by repeated integration be parts,
\begin{equation}
\begin{matrix}{
\int\limits_{z_1}^{z_2} dz f(z) e^{ikz} =
\sum\limits_{n=0}^{N-1} (-1)^n
\left[{f^{(n)}(z) e^{ikz} \over (ik)^{n+1}}\right]_{z_1}^{z_2} \cr
+ {(-1)^N \over (ik)^N} \int\limits_{z_1}^{z_2} dz f^{(N)}(z) e^{ikz}
}\end{matrix}
\end{equation}
which is valid provided the 0th through $N$th derivatives
are well defined. Thus, if $f(z_1) \ne f(z_2)$ then we
will find a leading order term $\tilde f \propto 1/k$, otherwise
we have $\tilde f \sim 1/k^2$, unless $f'(z_1) = f'(z_2)$
in which case $\tilde f \sim 1/k^3$ and so on.  

Here we have $z_1 = 0$, $z_2 \rightarrow \infty$, and we will
assume that the integrands and all necessary derivatives
vanish as $z \rightarrow \infty$, so provided $z n(z) \rightarrow 0$
and $n(z) / z$ is finite as $z \rightarrow 0$ then the
two fourier integrals in (\ref{eq:tildeg}) 
fall at least as fast as $1/k^3$ for large $k$.
Thus, the dominant term at high frequency is the first term 
$\tilde g = - 1/k^2$
and hence $|\tilde g|^2 \sim 1/ k^4$ for large $k$. 
The $\mu$-integral in
(\ref{eq:scalarGammaVariance}) is therefore dominated by modes with
$\mu \lsim 1/kD$ for which  $(1 - \mu^2)$
is very close to unity and consequently
\begin{equation}
\begin{matrix}{
\langle \Gamma_0^2 \rangle \simeq {1 \over 4 \pi^2}
\int\limits_0^\infty dk k^5 P(k) \int\limits_{-\infty}^\infty dy |\tilde g(y)|^2 \cr
\quad\quad\quad\quad
\sim D^3 \int dk k^5 P(k) \sim (D/\lambda)^3 \langle h^2 \rangle
}\end{matrix}
\end{equation}
which is the usual weak lensing result.

For the tensor case things are somewhat more complicated:
Now the leading order contribution to
$|\tilde g((1-\mu)k)|^2$ in (\ref{eq:gammacovariance}) falls
as $1/ ((1-\mu)k)^4$ for large $k$, but the pair of $R_+^2$ and
$R_\times^2$ terms contain 6 and 4 factors of $1-\mu$
respectively, so now
the $\mu$-integration is {\em not\/} restricted to small $1-\mu$
and has value $\sim 1/k^4$ and hence we find typical
covariance matrix elements $\langle \Gamma_{\alpha}^2 \rangle
\sim \langle h^2 \rangle$, with all four components of $\Gamma_\alpha$
having similar mean-square expectation values.  
However, the leading order term in $\tilde g$ is also
independent of the depth of the survey, so what we have
calculated here is clearly the shear due to the
local sky-distortion effect.  To extract the non-locally generated
shear we need to subtract from $\tilde g$ the distance independent
contribution.  We then find asymptotes $|\tilde g(k)|^2 \sim 1/k^4$ for
$k \ll 1/D$ and $|\tilde g(k)|^2 \sim 1/(D^2 k^6)$ for $k \gg 1/D$.
We now find that the dominant contribution to $\langle \Gamma_\alpha
\Gamma_\beta \rangle$ comes from the `$\times$'-polarization
and is dominated by the nearly resonant
modes with $1 - \mu \simeq 1/kD$.  Now from
(\ref{eq:Ralphadef}) one can see that $\Gamma_{\times 0} = 0$, so the 
$\times$-polarization gives no convergence;
$\Gamma_0 = 0$, so there is no area amplification (in accord with Zipoy and
Bertotti), and 
$ \langle \Gamma_1^2 \rangle =  \langle \Gamma_2^2 \rangle =
\half \langle \Gamma_3^2 \rangle$ with mean-square rotation
\begin{equation}
\label{eq:tensorGammasquared}
\langle \Gamma_3^2 \rangle \simeq {1 \over 8 \pi^2}
\int\limits_{0}^\infty dk k P(k) \int\limits_{-\infty}^\infty dy y^4 
| \tilde g(y) |^2 \sim (\lambda / D) \langle h^2 \rangle
\end{equation}
which is smaller than the scalar case by a factor $(\lambda / D)^4$.

It may seem a little surprising that $\langle \delta \theta^2 \rangle \sim
\langle h^2 \rangle (\lambda / D)^2 \ln(D/\lambda)$ yet the mean-square shear
$\langle \gamma^2 \rangle = 
\langle \Gamma_1^2 + \Gamma_2^2\rangle \sim (\lambda / D) \langle h^2 \rangle $,
since, on general grounds, one expects $\gamma \sim \delta\theta / \theta_c$
where $\theta_c $ is the coherence scale for the angular deflections
and so it would appear that $\theta_c \sim \sqrt{\lambda/D}$, which
is much less than $\lambda/D$ which is the angle subtended by the characteristic
wavelength or coherence scale for the metric fluctuations.
The reason for this is that the shear is dominated by
the nearly resonant modes with $1 - \mu \sim 1/kD$, which run
nearly parallel to the line of sight, and these have a projected
transverse wavelength $\sim \lambda / \sqrt{1 - \mu^2} \sim \sqrt{\lambda D}$,
and therefore subtend an angle $\theta_c \sim \sqrt{\lambda / D}$.

We have found that the leading order non-local effect gives
no convergence whatsoever. However, there is in fact some
flux modulation due to the Sachs Wolfe
effect, which causes a modulation in the surface brightness 
given by the difference in the metric fluctuations at the
source and observer.  One can calculate the mean square amplification
as before by summing over plane waves, and again only the nearly
resonant modes contribute and we find $\langle \Gamma_0^2 \rangle^{1/2}
\sim \sqrt{\lambda/D} h$ which is of the same order as the
shear and rotation.

The shear $\gamma \sim h$ due to the local sky distortion
is much larger than the spatially inhomogeneous shear $\gamma
\sim \sqrt{\lambda / D} h$ due to the metric fluctuations along the 
line of sight.  The gradients of these two contributions to the
shear are of similar order, however, and high frequency fluctuations in the
shear are dominated by the non-local term.  

We have assumed above that $n(z) \simeq \overline n(z)$.
Fluctuations in $n(z)$ can increase the predicted shear and
amplification somewhat.  If we consider the rather extreme case
of a narrow beam which intercepts only one cluster
(of width $\delta z_c$) then one finds that $\langle \gamma^2 \rangle
\sim \langle h^2 \rangle$ if $\delta z_c \ll 1/k$ and
$\langle \gamma^2 \rangle \rangle / (k \delta z_c)$ if
$\delta z_c \gg 1/k$, so for a cluster which is smaller than
the characteristic wavelength the shear, amplification etc.~can be
of order $h$.  However, this does not make the effect any easier
to observe since in the situation we have described --- which might
be a good model for a narrow `pencil-beam' survey --- the fluctuations
in the number density from beam to beam would be of order unity.

Our analysis has yielded a very different result from Linder's.
We have not been able to pinpoint the source of the discrepancy, but is would
seem to be a consequence of inappropriate boundary conditions;
one can view (\ref{eq:geoddev}) as a second order ODE for $\psi_{lm}$.
For generic initial conditions, one will find, in addition to the
particular integral solution terms which are constant or grow linearly
with source distance, and which could plausibly account for Linder's result.
However, as we have seen, with the correct boundary conditions at 
the observer these terms are absent.

The weak lensing effects obtained above are very small indeed.
For the local term, we find $\gamma$ etc.~on the order of
$\sim h \sim \sqrt{\OmegaGW} H_0 / \omega$, and for the
non-local term, and for distant sources
($D \sim 1/H_0$) we have $\gamma \sim \sqrt{\OmegaGW} (H_0 / \omega)^{3/2}
\sim \sqrt{\OmegaGW} \theta_c^{-3}$, and therefore for waves
with $\theta_c \sim 10^{-2}$ say we would find an induced clustering
effect $w(\theta \sim \theta_c) \sim 10^{-12} \OmegaGW$ which is very small for
any plausible value for $\OmegaGW$.

\section{Discussion}

We have calculated the statistical properties of photon
trajectories propagating through a statistically
homogeneous and isotropic sea of metric fluctuations.

Stimulated by the power of lensing phenomena to constrain
small amplitude metric perturbations and the theoretical
expectation that the dominant metric fluctuations on
small-scales 
are most likely predominantly in the form of gravity waves,
we have focussed on the effect of metric fluctuations with
wavelength much less than the path length.

We have found qualitatively different behavior
depending on whether the metric fluctuations are
scalar or tensor (gravity wave) fluctuations. 
In the former case, the photon trajectories can
be pictured as suffering a set of uncorrelated
kicks which give a random walk for the
photon momenta (and the displacement, which is the
integral of the transverse momentum, grows as
the $3/2$ power of path length).  For gravity waves we have
found that this picture of uncorrelated kicks is
grossly misleading and that the trajectories are
in fact a scale-invariant `flicker-' or `$1/f$-noise'
process. While a plot of a photon trajectory
over a short distance of a few gravity wavelengths would
look qualitatively similar for the two cases, on larger
scales the coupling for gravity waves is suppressed by
$(\lambda / D)^{3/2}$, and the photons can propagate over
great distances with barely any deflection.  

In addition to this rather general result, we have made detailed
calculations of a number of potentially observable
effects: proper motions of distant sources; lateral displacement
of caustic networks; and weak-lensing of distant
objects.  In all cases the estimated size of the
effects are too small to be of much practical interest
to experimentalists, counter to some claims in the
literature.  

Our analysis has concentrated exclusively on stochastic
backgrounds of gravity waves and so our results cannot
be directly applied to deflection of light passing
close to a binary system emitting gravity waves (as discussed
extensively by Fakir and Durrer).  We have however, attempted to
repeat these calculations. For very small
impact parameters ($b \lsim \lambda$) we confirm
the deflection on the order of $\delta \theta \sim
h(b)$ as found by Fakir (1994a,b) (though we note that
this effect is on the same order as the scalar metric
perturbation associated with the binary)
but that at larger impact parameters the effect
falls off as a high power of $1/b$, and we were not
able to confirm Durrer's claim that $\delta \theta \sim
h(b)$ in general.

To end on a slightly more positive note,
we have found that the dominant astrometric effect of
a gravity wave background is the local distortion of the
sky due to the metric fluctuation in the vicinity
of the observer (see Pyne \etal, 1996), which will cause
proper motions $\delta \dot \theta \sim \omega h$.
Barring the possibility of a strong background of high
frequency waves from coalescing binaries, the dominant
contribution to this effect would most likely come from
waves on the order of the present horizon size.  While
the predicted size of the motions is impressively small 
($\sim 10^{-9}$arcsec/yr), and would typically be similar in amplitude
to the $\sim 30$km/s or so motions due to horizon scale
scalar metric fluctuations they might in principle be measurable
as a relative motion of distant galaxies.

\section{References}

\noindent
Abbott, L., and Wise, M., 1984. Nuc.~Phys.~B., 244, 541-548

\noindent
Allen, B., 1996. preprint, gr-qc/9604033

\noindent
Bar-Kana, R., 1996. To appear in Phys.~Rev.~D., astro-ph/9606065

\noindent
Battye, R., Caldwell, R., and Shellard, E., 1996. preprint, astro-ph/9607130

\noindent
Bergmann, P., 1971. Phys.~Rev.~Lett., 26, 1398-1400

\noindent
Blandford, R., Saust, A., Brainerd, T., and Villumsen, J., 1991.
MNRAS, 251, 600

\noindent
Braginsky, V., Kardashev, N., Polnarev, A., and Novikov, I.,
1990. Il Nuovo Cimento, 105B, 1141-1158

\noindent
Durrer, R., 1994. Phys.~Rev.~Lett., 72, 3301-3304

\noindent
Fakir, R., 1992. ApJ, 418, 202

\noindent
Fakir, R., 1994a. ApJ, 426, 74

\noindent
Fakir, R., 1994b. preprint, gr-qc/9411042

\noindent
Gould, A., 1992. ApJ, 368, L5

\noindent
Kaiser, N., 1992. ApJ, 388, 272-286

\noindent
Linder, E., 1987. Ph.D. thesis, Stanford University

\noindent
Linder, E., 1988. ApJ, 328, 77-87

\noindent
Marleau, F., and Starkman, G., 1996. preprint, astro-ph/9605066

\noindent
Miralda-Escude, J., 1991. 1991, ApJ, 370, 1

\noindent
Paczynski, B., 1986. ApJ, 304, 1

\noindent
Press, W.H., 1978. Comments Astrophys, 7, 103

\noindent
Pyne, T., Gwinn, C., Birkinshaw, M., Marshall-Eubanks, T., and Matsakis, N., 
1995. preprint, astro-ph/9507030

\noindent
Zipoy, D., and Bertotti, B., 1968. Il Nuovo Cimento, 56B, 195-197

\end{document}